\documentclass[a4paper,10pt]{article}
\usepackage[utf8]{inputenc}
\usepackage[T1]{fontenc}
\usepackage{hyperref}
\usepackage{amsthm,amsfonts,amsmath,amssymb}
\usepackage{textcomp}
\usepackage{multirow}
\usepackage{multicol}
\usepackage[a4paper, top=2cm, bottom=2cm, left=2.5cm, right=1cm]{geometry}
\usepackage{tabularx}

\title{Application of data compression techniques to time series forecasting}
\author{K.S. Chirikhin$^{a, \ast}$, B.Ya. Ryabko$^{b, a, \ast \ast}$\\
\vtop{\small \it $^{a}$Novosibirsk State University, Novosibirsk, Russia}\\
\vtop{\small \it $^{b}$Institute of Computational Technologies, Siberian Branch of the Russian Academy of Sciences,
Novosibirsk, Russia}\\
\vtop{\small $\ast${\it email:} \url{chirihin@gmail.com}}\\
\vtop{\small $\ast\ast${\it email:} \url{boris@ryabko.net}}}

\begin{document}

\date{}
\maketitle

\begin{abstract}

It is known in Information Theory that the problems of data compression and time series forecasting are closely related. More precisely, in the case of stationary processes, a data compressor which gives an asymptotically minimal length of a compressed sequence can be used as a time-series predictor whose prediction error (per symbol) is asymptotically minimal, see details in the book (Ryabko, Astola, Malyutov, Compression-Based Methods of Statistical Analysis and Prediction of Time Series. Springer, 2016). In this study we show that standard well-known file compression programs (zlib, bzip2, etc.) are able to forecast real-world time series data well. The strength of our approach is its ability to use a set of data compression algorithms and "automatically" choose the best one of them during the process of forecasting. Besides, modern data-compressors are able to find many kinds of latent regularities using some methods of artificial intelligence (for example, some data-compressors are based on finding the smallest formal grammar that describes the time series). Thus, our approach makes it possible to apply some particular methods of artificial intelligence for time-series forecasting.

As examples of the application of the proposed method, we made forecasts for the monthly T-index and the Kp-index time series using standard compressors. In both cases, we used the Mean Absolute Error (MAE) as an accuracy measure.

For the monthly T-index time series, we made 18 forecasts beyond the available data for each month since January 2011 to July 2017. We show that, in comparison with the forecasts made by the Australian Bureau of Meteorology, our method more accurately predicts one value ahead.

The Kp-index time series consists of 3-hour values ranging from 0 to 9. For each day from February 4, 2018 to March 28, 2018, we made forecasts for 24 values ahead. We compared our forecasts with the forecasts made by the Space Weather Prediction Center (SWPC). The results showed that the accuracy of our method is similar to the accuracy of the SWPC's method. As in the previous case, we also obtained more accurate one-step forecasts.
\end{abstract}

\section{Introduction}
The problem of time series forecasting attracts attention of a vast number of researchers due to its great practical importance in prediction of various economic, social and physical phenomena. There are several approaches to that problem: autoregression models \cite{box}, exponential smoothing \cite{exponential}, neural networks \cite{neural, ann-ar} and many others. Despite this, the problem of improving the accuracy of forecasting remains relevant.

In this paper we develop an algorithm of time series forecasting based on data compression techniques. The detailed description of compression-based approach to time-series forecasting can be found in \cite{ryabko-coding}. Nowadays, there are many efficient lossless data-compressors (or archivers) which are widely used in information technologies. These compressors are based on different ideas and approaches, among which, we note the PPM universal code \cite{ppm}, Burrows-Wheeler transformation \cite{bwt}, dictionary-based compression algorithms \cite{lz77, lz78}, grammar-based codes \cite{grammar}. 

The main contribution of this study is that we show how to apply standard, well-known file compression programs to forecast real-world time series. It's important to note that modern data compression algorithms use a variety of heuristics to improve the compression ratio. Thus, our approach allows to use algorithms with proven efficiency. Moreover, our method is able to use a set of algorithms and "automatically" select the most accurate among them. Besides, the described approach gives a possibility to apply some methods of artificial intelligence for time-series forecasting. Those methods are used in modern data compression algorithms in order to find many kinds of latent regularities.

The rest of the paper is organized as follows. In section 1 we briefly describe the relationship between prediction of finite alphabet sequences and data compression, whereas the section 2 discusses the generalization of that approach to real-valued series. Section 3 presents results of our experimental investigation of the described method and the section 4 contains a conclusion.

\section{Forecasting of finite alphabet sequences}
We begin the description of the proposed algorithm with an explanation of the relationship between the problems of data compression and prediction of sequences. Denote a finite set of symbols (an alphabet) as \(\mathcal{A}\), and the set of all possible words (sequences) with length \(n\) over alphabet \(\mathcal{A}\) as \(\mathcal{A}^{n}\). A lossless code \(\phi\) is a mapping \(\phi: \mathcal{A}^{n} \rightarrow \{0,1\}^{\ast}\), such that for any sequence of words \(x_{1}, x_{2}, \ldots, x_{m} \in \mathcal{A}^{n}\), \(m > 1\), \(n \ge 1\) the sequence \(\phi(x_{1}), \phi(x_{2}), \ldots, \phi(x_{m})\) can be uniquely decoded as \(x_{1}, x_{2}, \ldots, x_{m}\). In the paper \cite{ryabko-coding} the following formula was proposed to obtain a probability distribution over \(\mathcal{A}^{n}\) using a data compression method \(\phi\):
\begin{equation}
\label{eq:Kraft}
P_{\phi}(X) = \dfrac{2^{-|\phi(X)|}}{\sum \limits_{Y \in \mathcal{A}^{n}} 2^{-|\phi(Y)|}},
\end{equation}
where \(X \in \mathcal{A}^{n}\), \(|\phi(X)|\) is the length of encoded representation of \(X\).

The estimation of the conditional probability that the next symbol \(x_{t+1}\) will be equal to \(a \in \mathcal{A}\) in a sequence \(x_{1},x_{2},\ldots,x_{t}\) can be obtained by the following formula:
\begin{equation}
\label{eq:distr}
P_{\phi}(x_{t+1} = a|x_{1}x_{2}\ldots{}x_{t}) = \dfrac{P_{\phi}(x_{1}x_{2}\ldots{}x_{t}a)}{\sum \limits_{b \in \mathcal{A}} P_{\phi}(x_{1}x_{2}\ldots{}x_{t}b)}.
\end{equation}

The strength of our approach is the ability to use a set of algorithms and "automatically" select the most accurate among them. Suppose that we have several data compression methods \(\phi_{1}, \phi_{2}, \ldots, \phi_{k}\) and each of them works well with a particular type of sequences. In this case, we can obtain a single method out of them which has almost the same accuracy as the most accurate one on each type of sequence. This can be done by using the following formula:

\begin{equation}
\label{eq-comb}
P_{\phi}(x_{t+1} = a|x_{1}x_{2}\ldots x_{t}) = \dfrac{\sum \limits_{i = 1}^{k}\omega_{i}2^{-|\phi_{i}(x_{1}x_{2}\ldots x_{t}a)|}}{\sum \limits_{b \in \mathcal{A}}\sum \limits_{i=1}^{k} \omega_{i}2^{-|\phi_{i}(x_{1}x_{2}\ldots x_{t}b)|}},
\end{equation}
where the sum of non-negative weight coefficients \(\omega_{i}\) is equal to \(1\).

We illustrate the workings of the algorithm using a simple example. Consider how to predict the next two values of the sequence \(X = 00011100011100011\) (according to the pattern they are \(10\)). The length of the sequence is 17 characters. We in turn add all possible sequences of length two from the alphabet \(\{0, 1\}\) at the end of \(X\). As a result, we obtain four sequences of length 19. Then we compress the "extended" sequences encoded in ASCII using the zlib and ppmd compressors. The results are shown in table \ref{example1}. In the second and third columns the sizes of compressed files are presented.
\begin{table}[ht]
	\centering
	\begin{tabular}{|c|c|c|c|}
		\hline Sequence \(X'\) & zlib, bits & ppmd, bits & $P_{\phi}(X')$ \\\hline
		$0001110001110001100$ & 144 & 120 & 0.0039 \\\hline
		$0001110001110001101$ & 144 & 120 & 0.0039 \\\hline
		$0001110001110001110$ & 128 & 112 & 0.9884 \\\hline
		$0001110001110001111$ & 136 & 120 & 0.0039 \\\hline
	\end{tabular}
	\caption{Probability estimates for all possible combinations of the next two values}
	\label{example1}
\end{table}

For instance, let's compute the probability of the "correct" next values \(10\) by the formula \ref{eq:distr}. We use the equal weights \(\omega_{1} = \omega_{2} = 0.5\):
\[P(10|X) = \dfrac{0.5 \cdot 2^{-128} + 0.5 \cdot 2^{-112}}{0.5 \cdot (2*2^{-144} + 2^{-128} + 2^{-136}) + 0.5 \cdot (3*2^{-120} + 2^{-112})} \approx 0.9884.\]

We can obtain the probability that the next symbol will be \(0\) summing up the probabilities of \(00\) and \(01\) (we got the sum of probabilities equals to \(1.0001\) due to rounding errors):
\[P_{\phi}(0|00011100011100011) = 0.0039 + 0.0039 = 0.0078;\]
\[P_{\phi}(1|00011100011100011) = 0.9884 + 0.0039 = 0.9923.\]

We can use the mean value \(0.9923\) as an one-step forecast.

\section{Forecasting of real-valued sequences}
The described algorithm can be used to forecast real-valued time series.
Suppose we have a time series \(x_{1}, x_{2}, \ldots, x_{t}\), \(x_{i} \in \mathbb{R}\) and we want to
predict the next \(h\) values \(x_{t+1}, x_{t+2}, \ldots, x_{t+h}\). The basic idea is to transform
the original real-valued series to the finite alphabet sequence. Denote as \([m, M]\) an interval containing all \(x_{i}\), \(1 \le i \le t\). We divide this interval into \(k\) disjoint subintervals of equal length with numbers \(0,1, \ldots, k-1\) (we denote a subinterval with number \(i\) as \(q_{i}\)). Then we can replace each value in the original time series \(x_{i}\) with the corresponding subinterval number \(q_{j}\) and obtain the new time series of subinterval numbers \(Y = y_{1}, y_{2}, \ldots, y_{t}\), \(y_{i} \in \{0,1, \ldots, k-1\}\). We make a forecast of subintervals containing the next \(h\) values of the series \(X\) using the series \(Y\). We can obtain the probability that at the time \(t + i\), \(1 \le i \le h \), the value of the series \(X\) will fall into the subinterval with number \(j\) from the marginal probability distribution of the subinterval numbers:

\begin{equation}
\label{eq:inf}
P_{\phi}(x_{t+i} \in q_{j}) = \sum \limits_{a_{1}\ldots,a_{i-1}a_{i+1},\ldots,a_{h}\in \mathcal{A}^{h-1}} P_{\phi}(a_{1}\ldots a_{i-1},j,a_{i+1},\ldots,a_{h})
\end{equation}

Next we consider the problem of selecting the number of subintervals \(k\). This parameter has a great influence on the accuracy of the method. If \(k\) is too small we may obtain low accuracy due to rounding. On the other hand, if \(k\) is too large, then the noise in the data can decrease the accuracy. Moreover, with the growth of \(k\) the computational complexity of the algorithm grows exponentially. In this paper we use the approach described in \cite{ryabko-nonparametric}. We partition an interval containing all values of a time series by \(2^{i}\) subintervals, \(i = 1,2,\ldots,n\), \(k = 2^{n}\). Then we make independently forecasts at each \(i\) and combine this forecasts with weight coefficients. Denote as \(x_{i}\) the term of the original time series and as \(y_{i}^{j}\) the interval number corresponding to it in the partition into \(2^{j}\) intervals. We can use all partitions with weight coefficients by formula \ref{eq:mult_int}:

\begin{equation}
\label{eq:mult_int}
P_{\phi}(y_{1}^{n}, y_{2}^{n},\ldots,y_{t}^{n}) = \dfrac{\sum \limits_{i=1}^{n}\omega_{i}2^{-|\phi(y_{1}^{n}, y_{2}^{n},\ldots,y_{t}^{n})| + t(n-i)}}{\sum \limits_{i=1}^{n}\sum \limits_{Z \in N_{i}^{t}} \omega_{i}2^{-|\phi(Z)| + t(n-i)}},
\end{equation}

where \(N_{i} = \{0,1,\ldots,2^{i}-1\}\) is the alphabet of subinterval numbers and the non-negative weights \(\omega_{i}\) sum to \(1\).

In the calculations by formula \ref{eq:mult_int} the probability distribution, obtained by the most compressible series, will dominate in the final result. Usually the series, obtained by partition into a smaller numbers of subintervals, are better compressible. For a more fair comparison of the lengths of codewords we add \(t(n-i)\) bits to the length of each word. Consider a simple example. Suppose that an interval that contains all values of the time series is partitioned into \(2\) and \(4\) subintervals (we denote them as partitions 1 and 2 correspondingly). Note that each subinterval in partition 1 corresponds to two subintervals in partition 2. To specify, in which subinterval of partition 2 falls a value from partition 1, a single bit is required (we can encode as 0 the lower half of the subinterval from partition 2 and as 1 the upper one). If there are \(t\) values in the time series, \(t\) bits are required for the whole series. And if the maximal partition has \(2^{n}\) subintervals, we must use \(n - i\) bits to specify to which subinterval of the maximal partition belongs a value from the partition into \(2^{i}\) subintervals.

\section{Experimental investigation}
In this section, we present the results of forecasting data from the real world using the described method. The experiments were performed on two time series: the T-index time series and the planetary K-index (Kp) time series.

We used three data compression programs to obtain all presented forecasts: zlib \footnote{\url{https://zlib.net}} (version 1.2.11), ppmd \footnote{\url{https://github.com/Shelwien/ppmd_sh}} (an implementation of Prediction by Partial Matching data compression algorithm) and grammar-based compressor Re-Pair \cite{rp}.

To evaluate the accuracy of a forecast we used Mean Absolute Error (MAE):
\[\text{MAE} = \dfrac{1}{h}\sum \limits_{i=1}^{h} |\hat{x}_{i} - x_{i}|,\]

where \(\hat{x}_{i}\) is the predicted value, and \(x_{i}\) is the observed value.

In our method the number of sequences to compress depends exponentially on the forecasting horizon. In order to reduce that number we used a procedure which can be shown by a simple example. Consider a series \(x_{1}x_{2}x_{3}x_{4}x_{5}x_{6}x_{7}\). To make a forecast for the next four values we can split that series into two series \(x_{1}x_{3}x_{5}x_{7}\) and \(x_{2}x_{4}x_{6}x_{8}\). Then, instead of making one forecast for four values ahead, we can make two forecasts, but just for two values ahead each. Using the first series we can predict \(x_{9}\) and \(x_{11}\), and using the second series we can predict \(x_{10}\) and \(x_{12}\). If we hadn't used that technique we would have to compress \(\mathcal{A}^{h}\) sequences. The described approach reduces that number to the \(2 \mathcal{A}^{h/2}\) sequences and can be obviously generalized.

We begin with the description of the monthly T-index time series forecasting. The T-index is an indicator of the highest frequencies able to be retracted from regions in the ionosphere \footnote{\url{https://www.sws.bom.gov.au/HF_Systems/1/6}}. A collection of data can be found on the web-site of the Space Weather Services (SWS) of Australian Bureau of Meteorology follow the link \url{http://listserver.ips.gov.au/mailman/listinfo/ips-tindex-predictions}. For each month since November 2000 a file containing all observed values since January 1938 until the previous month and forecasts for several years ahead is published. Our computations were carried out as follows. For each month since January 2011 to July 2017 we made our own forecasts for 18 values ahead using observed values until that month. Then, using the file 18 months later, we compared the MAE of our forecasts and the SWS forecasts. We used the following time series preprocessing techniques:

\begin{enumerate}
	\item To remove the seasonal component from the series we used Seasonal Trend Decomposition (STL) \cite{stl}. The frequency of the seasonal component was equal to 11 years or 132 month (because the length of a solar cycle is typically 10 to 11 years);
	\item We used smoothing function: \(x_{t}^{\ast} = \dfrac{2x_{t} + x_{t-1} + x_{t-2}}{4}\);
	\item We split a time series to the 6 time series as described previously (so instead of making one forecast for 18 values ahead, we made 6 forecasts for 3 values);
	\item We took a first difference, i.e. instead of forecasting a series \(x_{1},x_{2},\ldots,x_{t}\) we considered the series \(x_{2}-x_{1}, x_{3} - x_{2}, \ldots,x_{t} - x_{t-1}\);
	\item Despite the fact that all values in this time series are integer, we considered it as a real-valued sequence and used the technique of partition to subintervals in order to reduce the size of the alphabet. The maximal number of dubintervals were 16 (so we considered partitions to 2, 4, 8 and 16 intervals). But when computing MAE, we rounded our forecasts to integers.
\end{enumerate}

The results are presented in table \ref{table1}. As we can see from table \ref{table1}, our method is more accurate when forecasting one value ahead.
\begin{table}
	\label{table1}
	\centering
	\begin{tabularx}{\textwidth}{|p{1.5cm}|X|X|X|X|X|X|X|X|X|X|X|X|X|}
		\hline
		\multirow{2}{*}{} & \multicolumn{10}{c|}{MAE for forecasting horizon} & \multicolumn{3}{c|}{Average}\\
		\cline{2-14}
		& 1 & 2 & 3 & 4 & 5 & 6 & 8 & 12 & 15 & 18 & \mbox{1-4} & \mbox{1-8} & \mbox{1-18} \\
		\hline
		SWS forecast & 12.2 & 13.4 & 14.5 & 15.6 & 16.3 & 17.7 & 20.1 & 21.5 & 22.8 & 24.1 & 13.9 & 16.1 & 19.5 \\
		\hline
		zlib & 11.3 & 13.9 & 15.8 & 16.3 & 15.6 & 18.6 & 20.4 & 26.8 & 27.0 & 38.1 & 14.3 & 16.21 & 20.8 \\
		\hline
		ppmd & 11.2 & 14.2 & 16.7 & 17.9 & 17.2 & 19.0 & 24.2 & 26.9 & 31.5 & 28.3 & 15.0 & 18.0 & 22.8 \\
		\hline
		rp & 13.5 & 17.0 & 22.4 & 26.9 & 23.8 & 17.3 & 28.8 & 24.7 & 33.5 & 36.8 & 20.0 & 22.6 & 28.9 \\
		\hline
		zlib + ppmd + rp & 11.2 & 14.2 & 16.7 & 17.9 & 17.3 & 19.0 & 26.3 & 26.8 & 31.6 & 28.3 & 15.0 & 18.3 & 23.1 \\
		\hline
	\end{tabularx}
	\caption{The results of the T-index time series forecasting}
\end{table}

Let us proceed to the description of Kp-index time series forecasting. The planetary K-index is used to characterize the magnitude of geomagnetic storms \footnote{\url{https://www.swpc.noaa.gov/products/planetary-k-index}}. It's an integer value from range \(0, \ldots, 9\). The Space Weather Prediction Center (SWPC) \url{https://www.swpc.noaa.gov/products/planetary-k-index} publishes 3-hour K-index data (8 values per day) and forecasts for three days ahead. We made 24-steps forecasts for each day from 4 February 2018 to 28 March 2018 and compared the accuracy of our predictions with the accuracy of predictions made by SWPC. We used the following time series preprocessing techniques:

\begin{enumerate}
	\item We split a time series to the 8 time series as described previously (so instead of making forecasts for 24 values ahead, we made 8 forecasts for 3 values ahead);
\end{enumerate}

The results are summarized in table \ref{table2}.

\begin{table}
	\label{table2}
	\centering
	\begin{tabularx}{\textwidth}{|p{1.5cm}|X|X|X|X|X|X|X|X|X|X|X|X|X|X|X|}
		\hline
		\multirow{2}{*}{} & \multicolumn{11}{c|}{MAE for forecasting horizon} &  \multicolumn{4}{c|}{Average}\\
		\cline{2-16}
		& 1 & 2 & 3 & 4 & 5 & 6 & 8 & 12 & 15 & 18 & 24 & \mbox{1-4} & \mbox{1-8} & \mbox{1-18} & \mbox{1-24}\\
		\hline
		SWPC forecast & 1.11 & 1.26 & 0.98 & 1.06 & 1.09 & 1.11 & 1.30 & 1.11 & 1.25 & 1.40 & 1.21 & 1.10 & 1.16 & 1.16 & 1.13 \\
		\hline
		zlib & 0.70 & 1.00 & 1.15 & 0.91 & 1.00 & 1.09 & 1.10 & 1.02 & 1.25 & 1.57 & 1.45 & 0.94 & 1.00 & 1.16 & 1.19 \\
		\hline
		ppmd & 0.77 & 1.11 & 1.08 & 0.91 & 1.08 & 1.15 & 1.09 & 1.00 & 1.15 & 1.45 & 1.45 & 0.97 & 1.05 & 1.13 & 1.14 \\
		\hline
		rp & 2.04 & 2.06 & 2.28 & 1.83 & 2.02 & 2.08 & 2.06 & 1.58 & 1.70 & 2.62 & 2.57 & 2.05 & 2.09 & 2.04 & 2.09 \\
		\hline
		zlib + ppmd + rp & 0.77 & 1.11 & 1.08 & 0.91 & 1.08 & 1.15 & 1.09 & 1.00 & 1.15 & 1.45 & 1.45 & 0.97 & 1.04 & 1.27 & 1.16 \\
		\hline
	\end{tabularx}
	\caption{The results of the Kp-index time series forecasting}
\end{table}

As can be seen from the table \ref{table2}, the overall accuracy of the forecasts is very similar. As in the previous case, our method is more accurate when forecasting one value ahead.

\section{Conclusion}
In this paper we described how well-known data-compression programs can be used to forecast time series. We showed that such technique is competitive with widely-used methods and, in our opinion, can be used in practice.

\section*{Acknowledgment}
This work was supported by Russian Foundation for Basic Research (grant 18-29-03005	).

\end{document}